\documentclass[a4paper]{jpconf}
\usepackage{graphicx}
\usepackage{amsfonts,amsmath,amssymb,mathrsfs}

\def\be{\begin{equation}}
\def\ee{\end{equation}}
\def\bea{\begin{eqnarray}}
\def\eea{\end{eqnarray}}

\def\a{\alpha}

\def\m{\mu}
\def\n{\nu}

\def\a{\alpha}
\def\b{\beta}

\begin{document}
\title{Ho\v{r}ava--Lifshitz gravity: a status report}

\author{Thomas P. Sotiriou}

\address{Department of Applied Mathematics and Theoretical Physics, Centre for Mathematical Sciences, University of Cambridge, Wilberforce Road, Cambridge, CB3 0WA, UK}

\ead{T.Sotiriou@damtp.cam.ac.uk}

\begin{abstract}
This is intended to be a brief introduction and overview of Ho\v{r}ava-Lifshitz gravity. The motivation and all of the various versions of the theory (to date) are presented. The dynamics of the theory are discussed in some detail, with a focus on low energy viability and consistency, as these have been the issues that attracted most of the attention in the literature so far. Other properties of the theory and developments within its framework are also covered, such as: its relation to Einstein-aether theory, cosmology, and future perspectives.
\end{abstract}

\section{Introduction}
\subsection{Gravity and renormalizability}

It is a well known fact that general relativity is not a renormalizable theory. This means that, successful as it may be as a classical theory of gravity, it should be viewed as an effective theory that breaks down at some scale. Beyond that scale general relativity is not enough to describe the gravitational interaction or spacetime itself (depending on the perspective) and one cannot construct its quantum counterpart using conventional quantization techniques.

On the other hand, given that within this perspective general relativity is an effective theory, the Einstein-Hilbert action contains only the lowest order terms in a curvature expansion. A reasonable question is then whether a renormalizable theory can be obtained by including higher order curvature terms. Intuitively this could work, as such terms would modify the propagator of the graviton at high energies. Indeed, in 1977 Stelle showed that theories whose action includes invariants quadratic in the curvature are renormalizable \cite{stelle}. However, unfortunately this comes at a very high price, as such theories contain ghost degrees of freedom and are, therefore, not unitary. This is because by adding higher order curvature invariants in the action, one adds higher order time derivatives.

One could attempt to modify the propagator by adding higher order spatial derivatives without adding higher order time derivatives. This could presumably lead to a theory with improved ultraviolet (UV) behaviour without having to face the dreadful consequences of having higher order time derivatives. Obviously, following such a prescription requires treating spatial and time derivatives on a different footing and as such is in clash with Lorentz invariance. On the other hand, since the modified behaviour of the propagator is strictly needed in the UV, it is conceivable that Lorentz invariance could be recovered in the infrared (IR), or at least that Lorentz violations in the IR could stay below current experimental constraints.

Ho\v{r}ava's proposal \cite{Horava:2008ih,Horava:2009uw}, which has led to a spree of publications recently and is now commonly referred to as Ho\v{r}ava--Lifshitz gravity, can be seen as an attempt to put these heuristic arguments in a rigorous framework where one can test whether they are indeed valid when it comes to gravity theories. In what comes next, an overview of this framework is given and its major developments to date are briefly discussed. This is far from an exhaustive account of an already very extensive literature, but hopefully it covers the most critical topics and provides a concise introduction to the subject.\footnote{See also Ref.~\cite{Weinfurtner:2010hz} for an early review of the projectable Ho\v{r}ava--Lifshitz gravity.}

\subsection{Lorentz violations as a field theory regulator}

What has been discussed above is essentially the idea of giving up Lorentz invariance in order to keep a quantum field theory finite. Clearly, this idea goes beyond gravity theories and it has been considered in the past for other fields, see for example Ref.~\cite{anselmi}. In fact, it would be best to start from a simpler example, such as a scalar field. This section will closely follow Ref.~\cite{Visser:2009fg}.

Let us first of all point out that there is nothing wrong  with using Lorentz violations as a field theory regulator from a perspective of logical consistency, as there is no reason why a theory should necessarily exhibit Lorentz symmetry in the far UV. On the other hand, Lorentz violations are severely constrained in a wide range of energies and especially in the IR. Therefore, as mentioned before, the real question is whether one can construct a field theory that exhibits Lorentz violations which in the far UV lead to renormalizability, but remain suppressed below experimental accuracy at lower energies. This is difficult to answer without concrete models and enough to motivate further investigation.

Consider the scalar field action 
\be
S_\phi=\int dt dx^d\left\{ \dot{\phi}^2-\sum_{m=1}^z a_m\phi(-\Delta)^m \phi+\sum_{n=1}^M g_n \phi^n \right\},
\ee
where $\dot{}\equiv \partial/\partial t$, $\Delta=\vec{\nabla}^2$ is the spatial Laplacian and $z$ and $M$ are positive integers still to be specified.  A theory is said to be ``power counting renormalizable'' when all of its interaction terms scale like energy to some non-positive power, as in this case Feynman diagrams are expected to be convergent (or have at most a logarithmic divergence). To check whether this is true or not we first have to choose the engineering dimensions of space and time. The engineering dimensions
\be
[dt]=[\kappa]^{-z},\qquad [dx]=[\kappa]^{-1},
\ee
where $\kappa$ is a place holder symbol with dimensions of momentum, are compatible with the choice $a_z=1$ and the requirement that the action has to be dimensionless. This choice of engineering dimensions, therefore, appears to be natural in the UV. The dimensions for the scalar field then turn out to be
\be
[\phi]=[\kappa]^{(d-z)/2}.
\ee
It remains to check the scaling for the rest of the terms. Requiring that an interaction term scales like energy to some non-positive power is equivalent to the requirement that the coupling of this interaction scales like energy to some non-negative power (since the action is dimensionless). Also, we find it convenient to work in terms of momenta instead of energies. So, our theory will be power counting renormalizable if all couplings have non-negative momentum dimensions.

It is straightforward to verify that
\be
[a_m]=[\kappa]^{2(z-m)}, \qquad [g_n]=[\kappa]^{d+z-n(d-z)/2}.
\ee
It is then obvious that $a_m$ has non negative momentum dimension for all values of $m$. It is also easy to realize that, for $z\geq d$, $g_n$ has non-negative momentum dimensions for all values of $n$. When $z<d$, $g_n$ has non-negative momentum dimension only when $n\leq[2(d+z)]/(d-z)$. One can then easily recover well-known results. For example one can verify that the case $d=3$, $z=1$, $M=4$ which corresponds to the usual relativistic $\phi^4$ theory in $3+1$ dimensions is a power counting renormalizable theory. The $d=z=1$ case, {\em i.e.}~the relativistic scalar in $1+1$ dimensions, is power counting renormalizable for all values of $M$. In fact this case belongs to the class $z=d$ and is directly analogous to the case $d=z=3$, which will also have the same property and is of interest to us. In fact one can argue that any theory with $z=d$ is actually finite with normal ordering, while theories with $z>d$ are finite even without normal ordering, see Ref.~\cite{Visser:2009fg} for a more detailed discussion using the ``superficial degree of divergence''.

How would things go if instead of a scalar field we wanted to consider a graviton? The main difference lies in the graviton self-interaction vertices. In the case of the scalar field considered above, momenta did not enter these interactions. For a graviton, however, one would have to deal with self-interaction vertices which contain spatial derivatives \cite{Visser:2009ys}. This introduces some further complication, but it is not enough to spoil the power-counting renormalizability, as long as $z\geq d$, {\em i.e.}~as long as the action contains operators with at least $2d$ spatial derivatives. The presence of self-interaction vertices which contain spatial derivatives is enough though to argue that, unlike the scalar theory consider above, its graviton analogue will not be finite (even with normal ordering) \cite{Visser:2009ys}.

All of the arguments presented here support the statement that field theories that contain at least $2d$ spatial derivatives in $d+1$ dimensions are power counting renormalizable. Power counting renormalizability is a strong indication that a theory is indeed renormalizable. Little progress has been made so far in the case of gravity to check renormalizability beyond power counting for such theories. 

\section{From the Lifshitz scalar to non-relativistic gravity}

We proceed to consider the exact structure of a gravity theory with the characteristics mentioned in the previous section. We will work in $3$ dimensions from now on but generalizing to more than 3 dimensions is straightforward. The basic requirement is that the theory should have only two time derivatives but at least $2d$ spatial derivatives, that is $6$ in our case. The fact that we want to have more spatial than time derivatives implies that we need to treat space and time on different footing. This is naturally achieved by working in the Arnowitt--Deser--Misner (ADM) decomposition of spacetime
\begin{equation}
\label{admmetric}
d s^2 = - N^2 c^2 d t^2 + g_{ij}(d x^i + N^i d t) (d x^j + N^j d t).
\end{equation}
One is essentially forced to pick a preferred foliation of spacetime in order to write down the action according to the prescription laid out in the previous section. Consequently, such an action cannot be invariant under standard diffeomorphisms, as general relativity is. It can, however, be invariant under a more restricted set: foliation preserving diffeomorphisms, {\em i.e.}~space-independent time reparametrization together with time-dependent spatial diffeomorphisms 
\be
t\rightarrow \tilde{t}(t),\qquad x^i\rightarrow \tilde{x}^i(t,x^i).
\ee
This is the symmetry that the action will have to respect.

It is straightforward to realize that the only covariant quantity under spatial diffeomorphisms that contains a time derivative of the spatial metric is the extrinsic curvature
\begin{equation}
K_{ij} = {1\over2N} \left\{  \dot g_{ij} - \nabla_i N_j - \nabla_j N_i \right\},
\end{equation}
which also transforms like a scalar under time reparametrization. A dot denotes differentiation with respect to the time coordinate and $\nabla_i$ is the covariant derivative associated with the spatial metric $g_{ij}$. We want the theory to be second order in time derivatives, so we need to consider terms quadratic in the extrinsic curvature. Note that there are no invariants under the symmetry mentioned above that one can construct with time derivatives of the lapse $N$ or the shift $N_j$, without including higher order time derivatives of $g_{ij}$ as well. Therefore, the most general action one can write is
\begin{equation}
S=\frac{M_{\rm pl}^2}{2}\int d^3x d t N \sqrt{g} \left\{ K^{ij} K_{ij} - \lambda K^2 -V(g_{ij},N)\right\}\, ,
\end{equation}
where $M_{\rm pl}$ is a constant which, with some foresight, we identify with the Planck mass, $g$ is the determinant of the spatial metric $g_{ij}$, and $\lambda$ is a dimensionless running coupling. $V$ generically can depend on $g_{ij}$ and $N$ and their spatial derivatives. It does not contain time derivatives, neither does it depend on the shift $N_i$, as the symmetry of the theory does not allow one to construct any suitable invariants. Finally, power counting renormalizability requires $V$ to contain terms which are at least sixth order in spatial derivatives. From now on we will restrict ourselves to theories that contain sixth order time derivatives but not higher, as these are the simplest ones which satisfy the power counting renormalizability requirement.

\section{Potential, constraints and the various versions of the theory}

Clearly there are numerous terms one could include in $V$. Different choices have led to different versions of the theory. In this section we list these versions, briefly presenting their basic characteristics and differences. We refrain from discussing or commenting upon their viability or consistency, as this, together with their phenomenology, will be the subject of the coming section.

\subsection{Detailed balance}
Ho\v{r}ava proposed a symmetry that $V$ would have to satisfy, that would drastically reduce the number of invariants one should consider \cite{Horava:2009uw}. This symmetry is dubbed {\em detailed balance} and it is inspired by condensed matter systems. It sums up to the requirement that $V$ should be derivable from a superpotential $W$:
\be
V=E^{ij}{\cal G}_{ijkl}E^{kl},
\ee
where
\be
E^{ij}=\frac{1}{\sqrt{g}}\frac{\delta W}{\delta g_{ij}},
\ee
and 
\be
{\cal G}^{ijkl}=\frac{1}{2}\left(g^{ik}g^{jl}+g^{il}g^{jk}\right)-\lambda g^{ij}g^{kl}.
\ee
The most general action one can write with $V$ satisfying the conditions above is
\bea
\label{dbaction}
S_{db}=\frac{M_{\rm pl}^2}{2}\int d^3x d t N \sqrt{g} \Bigg\{ &&\!\!\!\!\!\!\!\!\!\!K^{ij} K_{ij} - \lambda K^2-\frac{\alpha^4}{M_{\rm pl}^4} C_{ij} C^{ij}+\frac{2\alpha^2\beta}{M_{\rm pl}^3}\frac{\epsilon^{ijk}}{\sqrt{g}}R_{il}\nabla_j R^l_{\phantom{a}k}-\frac{\beta^2}{M_{\rm pl}^2}R^{ij}R_{ij}\nonumber\\
&&+\frac{\beta^2}{4}\frac{1-4\lambda}{1-3 \lambda} R^2+\frac{\beta^2 \,\zeta}{1-3\lambda} R-\frac{3\beta^2 \, \zeta^2}{1-3\lambda} M^2_{\rm pl}\Bigg\}\, ,
\eea
where $\epsilon^{ijk}$ is the Levi-Civita symbol, 
\be
C^{ij}=\frac{\epsilon^{ikl}}{\sqrt{g}}\nabla^k\left(R^j_{\phantom{a}l}-\frac{1}{4} R \delta^j_{\phantom{a}l}\right)
\ee
is the Cotton tensor, which in 3 dimensions plays the role of the Weyl tensor, and $\alpha$, $\beta$ and $\zeta$ are dimensionless couplings. Notice that there are only 3 new couplings for a total of 6 terms in $V$.

The advantages of detailed balance is that it drastically reduces the number of terms one needs to consider and that it introduces a superpotential, which might simplify quantization. On the other hand, the action above contains a term that violates parity (the fifth order operator).\footnote{Parity violating terms, within or outside the framework of detailed balance, have been shown to lead to chiral primordial gravitational waves \cite{Takahashi:2009wc}.} Additionally, as is obvious, the last term in the action, which plays the role of a cosmological constant, is restricted to have the wrong sign with respect to the one suggested by observations when $\lambda>1/3$, as pointed out in Refs.~\cite{Sotiriou:2009gy,Sotiriou:2009bx} ($\lambda$ is supposed to be close to unity, if severe Lorentz violation are not to occur).\footnote{This is a bare cosmological constant. Also, one could analytically continue the parameters as explained in Ref.~\cite{Lu:2009em}, but in this case $W$ would cease to be real.} Note that there is nothing fundamental about detailed balance, so it is safe to consider it just a simplicity assumption.

\subsection{Projectable Ho\v{r}ava--Lifshitz gravity}

Another possible simplifying restriction proposed by Ho\v{r}ava in Ref.~\cite{Horava:2009uw} is that of {\em projectability}, which is the assumption that the lapse is just a function of time. That is $N=N(t)$. Again there is no fundamental principle behind such an assumption. Ho\v{r}ava's main motivation for initially considering it was that under this assumption one has enough gauge freedom to set $N=1$, as in general relativity. This is not possible without projectability, as the symmetries of the action allow only space independent time reparametrizations, as explained earlier. 

In any case, projectability drastically reduces the number of invariants $V$ can include even if detailed balance is not imposed, as all terms with spatial derivatives of $N$ would now vanish. Let us see that in more detail, following the line of Refs.~\cite{Sotiriou:2009gy,Sotiriou:2009bx}. Since $N$ is now just a function of time there is no invariant one can built out of it without using time derivatives. So $V$ will only depend on the metric and its spatial derivatives. Essentially, this means that the action should include all of the curvature invariants one can construct with $g_{ij}$ with up to six spatial derivatives. 

An important simplification comes from the fact that we are in 3 spatial dimensions. Therefore, the Weyl tensor vanishes identically and the Riemann tensor can be expressed in terms of the Ricci tensor. Taking Bianchi identities into account as well and ignoring surface terms, one can arrive at the conclusion that the most general action is
\bea
\label{paction}
S_p&=&\frac{M_{\rm pl}^2}{2}\int d^3x d t N \sqrt{g} \Bigg\{ K^{ij} K_{ij} - \lambda K^2 -g_0 \, M_{\rm pl}^2 -g_1 R -     g_2 \,M_{\rm pl}^{-2}\,R^2 -  g_3 \, M_{\rm pl}^{-2}\, R_{ij} R^{ij} \nonumber\\
&&\qquad\qquad\qquad\qquad- g_4 \,  M_{\rm pl}^{-4}\,R^3 - g_5 \,M_{\rm pl}^{-4}\, R (R_{ij} R^{ij})- g_6 \,M_{\rm pl}^{-4}\, R^i{}_j R^j{}_k R^k{}_i \nonumber\\
&&\qquad\qquad\qquad\qquad
- g_7\,M_{\rm pl}^{-4}\, R \nabla^2 R - g_8 \,M_{\rm pl}^{-4}\, \nabla_i R_{jk} \, \nabla^i R^{jk}\Bigg\}\, ,
\eea
where the $g_i$ are dimensionless couplings. As long as we have not coupled matter to the theory we are free to rescale the coordinate in order to set $g_1=-1$, which is the value it has in general relativity. We then have 8 remaining couplings $g_i$ which can be used to tune the scales suppressing the various operators. The following remarks are in order:
\begin{itemize}
\item We have suppressed parity violating terms.
\item Enforcing projectability on top of detailed balance (without suppressing parity violating terms) would leave action (\ref{dbaction}) unaffected, apart from the fact that $N$ would become just a function of time. As we will see later, this is quite an important difference.
\item There are just 3 more operators in the most general projectable action than in the one with detailed balance. In this sense, detailed balance does not bring significant simplification once projectability has been assumed. For this reason, and for the drawbacks listed in the end of the previous section, we will not consider detailed balance further when assuming projectabiltity.
\item $g_0$ controls the value of the cosmological constant, which, unlike the case with detailed balance, is not restricted.
\item There are two types of Lorentz violating terms in the action. The ones in $V$, which are suppressed by some scale that can be determined by tuning the couplings $g_2$ to $g_8$, and one that comes from the kinetic part due to the fact that $\lambda$ is not necessarily equal to 1. This will be a generic feature of all versions of the theory as we will see. Clearly, the term of the second kind is far more dangerous, as it introduces Lorentz violations at low energy scales.
\item All couplings are running. One can, therefore, hope that $\lambda$ will run to 1 in the IR (or sufficiently close to it for experimental constraints to be satisfied), the rest of the operators will be heavily suppressed, and the theory will effectively reduce to general relativity. Diffeomorphism invariance and Lorentz invariance would emerge as IR (approximate) symmetries.  Whether or not this will indeed be the case requires a study of the renormalization group flow which is still pending.
\end{itemize}

\subsection{Non-projectable Ho\v{r}ava-Lifshitz gravity}

We now proceed to consider the case where neither detailed balance nor projectability are enforced. This version of the theory is called non-projectable. To avoid confusion it is worth pointing out explicitly, that the version with detailed balance can be a non-projectable version. However, in this section we wish to go beyond that. Once one abandons detailed balance, adding just some specific choice of extra terms is not really an option. Radiative corrections will generate all possible terms compatible with the symmetries of the theory and, thus, all such terms should be taken into account, much like the projectable case treated above. However, as first pointed out in Ref.~\cite{Blas:2009qj}, it is not only curvature invariants of $g_{ij}$ that one can include in $V$ in this case. One can also use the quantity
\be
a_i=\partial_i \ln{N}\,,
\ee
as contractions of it with itself or curvature terms also lead to invariants. The lowest order invariant one can construct with $a_i$ is $a^i a_i$, which comes at the same order as $R$. The action will then be of the form
\bea
\label{npaction}
S_{np}&=&\frac{M_{\rm pl}^2}{2}\int d^3x d t N \sqrt{g} \Bigg\{ K^{ij} K_{ij} - \lambda K^2 +\xi R + \eta \, a_ia^i+\frac{1}{M_A^2}L_4+\frac{1}{M_B^4}L_6\Bigg\}\, ,
\eea
where $L_4$ and $L_6$  include all possible 4th and 6th order operators respectively that one can construct using $a_i$ and $g_{ij}$. For example, $R^2$, $(a^ia_i)^2$, $\nabla^i\nabla_i R$ and $a_ia_j R^{ij}$ are some of the terms that are included in $L_4$.\footnote{$\nabla^i\nabla_i R$ and other similar terms are not surface terms, as the $N$ in the 4-volume is space dependent.} As was the case before for $g_1$, $\xi$ can now be set to unit by a coordinate rescaling when matter is not coupled to the theory. This version of the theory was first studied in Ref.~\cite{Blas:2009qj} in order to resolve some inconsistencies in the dynamics of the non-projectable version with detailed balance, as we will discuss shortly. Although it is often referred to as an ``extension" of Ho\v{r}ava-Lifshitz gravity, this terminology is probably unfortunate: this is Ho\v{r}ava-Lifshitz gravity, {\em i.e.}~a theory constructed consistently according to the prescription described in Ho\v{r}ava's paper Ref.~\cite{Horava:2009uw} and without any restriction, such as detailed balance or projectability. 

The following points are worth stressing for the non-projectable version (apart from some of the remarks already made for the projectable case and apply here as well).
\begin{itemize}
\item For simplicity the cosmological constant term has been suppressed, but it can be straightforwardly restored.
\item The scales $M_A$ and $M_B$ suppressing the higher order operators have been left arbitrary. This is in analogy with having arbitrary dimensionless couplings $g_i$ in the projectable case. 
\item The number of operators in the non-projectable case is an order of magnitude larger than that in the projectable case.
\item If general relativity were to be recovered in the IR, not only $\lambda$ would have to run to 1 but also $\eta$ would have to run to 0.
\end{itemize}

\subsection{Imposing further symmetries}

All of the actions presented above, which correspond to different versions of Ho\v{r}ava--Lifshitz gravity, are invariant under foliation preserving diffeomorphisms only, and not the full set of diffeomorphisms. So, even though the action is in all cases quadratic in the time derivatives of $g_{ij}$, less symmetry generically implies that more degrees of freedom will be excited. We will see explicitly in the next section that this is indeed the case, and that this can, in most cases, have undesirable consequences for the consistency and viability of the theory.

A way to do away with the extra degrees of freedom is to attempt to nontrivially extend the gauge symmetry of the theory, so that it will have as many generators per spacetime point as general relativity. It was pointed out in Ref.~\cite{Horava:2010zj} that this could be done if the action would be suitably modified to acquire an extra local $U(1)$ symmetry. More precisely in Ref.~\cite{Horava:2010zj} a step by step construction of the action led to
\bea
\label{hmt}
S_{{\rm extra}\, U(1)}&=&\frac{M_{\rm pl}^2}{2}\int d^3x d t N \sqrt{g} \Big\{ K^{ij} K_{ij} - K^2 -V(g_{ij},N)\nonumber\\&&\qquad\qquad+\nu \Theta^{ij}(2 K_{ij}+\nabla_i\nabla_j \nu)-A(R-2 \Omega)/N\Big\}\, ,
\eea
where $\Omega$ is a constant, $A$ acts as a Lagrange multiplier, $\nu$ is an auxiliary scalar field,
\be
\Theta^{ij}\equiv R^{ij}-\frac{1}{2} g^{ij} R+\Omega g^{ij},
\ee
and $V$ should as usual contain 6th order operators and may or may not satisfy detailed balance. Note that $A$ transforms as a spatial scalar and a time vector under foliation preserving diffeomorpisms, that is
\be
A \rightarrow A+ \dot{f} A+f \dot{A}+\xi^i\partial_i A,
\ee
where $\xi^i$ is the generator. On the other hand, under a local $U(1)$ gauge transformation with generator $\varphi$, $A$ and $\nu$ transform as
\bea
&&A\rightarrow A+\dot{\varphi}-N^i\nabla_i \varphi,\\
&& \nu\rightarrow \nu+\varphi.
\eea

See Ref.~\cite{Horava:2010zj} for more details on constructing the action, as well as for discussions on anisotropic scaling, hamiltonian formulation, linearization, etc. It was argued there that in the IR this theory has very good chances to reduce to general relativity, as the high order operators in $V$ would be suppressed (as in the previous versions of the theory), the extra fields  in the action are just auxiliary fields, and the extra gauge symmetry leads to just the usual spin 2 graviton excitation. Moreover, it was claimed in Ref.~\cite{Horava:2010zj} that the $U(1)$ symmetry forces the coefficient $\lambda$ of $K^2$ in the action, which was a running coupling in all other versions of the theory, to be equal to 1 here, as in general relativity.

This last claim has been contested recently in Ref.~\cite{daSilva:2010bm} where it has been claimed that the extra $U(1)$ symmetry does not necessarily require $\lambda=1$ and that even for arbitrary $\lambda$ the theory propagates only a spin 2 graviton. If this is true, $\lambda$ has to remain a running coupling and flow close to $1$ in the IR for general relativistic phenomenology to be recovered according to Ref.~\cite{daSilva:2010bm}. So, at the moment, whether or not the value of $\lambda$ is fixed to 1 can be considered a matter of debate.

Another potential difficulty with this version of Horava-Lifshitz gravity has to do with matter coupling. It is straightforward to notice that the Lagrange multiplier in action (\ref{hmt}) forces $R$ to be constant. Even though this is not necessarily a problem in vacuum, it cannot be generically the case once matter is coupled to the theory. This implies that matter would have to be coupled to $A$ if the theory is to resemble general relativity at low energies, and that a suitable coupling that leads to standard phenomenology in the IR should exist. Whether this is indeed the case requires further investigation (some first steps where done in Ref.~\cite{daSilva:2010bm}). 
It is also worth pointing out that action (\ref{hmt}) is susceptible to radiative corrections which could modify its form.

Since this last version of Horava-Lifshitz gravity has been proposed very recently and has not yet been studied extensively, we will not consider it further. Its cosmological aspects where recently studied in Ref.~\cite{Wang:2010wi}.

\section{Projectable version: Dynamics, consistency and low energy behaviour}
\subsection{Degrees of freedom and propagators}

We return now to the projectable version of the theory and take a closer look at its dynamics. We will consider the most general action in this version, action (\ref{paction}), {\em i.e.}~we will not enforce detailed balance but our results will include it as a subcase. The field equations for this action have been derived in detail in Ref.~\cite{Sotiriou:2009bx}. We will refrain from rewriting them here due to space limitations but we will make the following remark: since $N=N(t)$, the variation with respect to $N$ will not lead to a usual local super-Hamiltonian constraint, but to a global Hamiltonian constraint. That is, the constraint will come in the form of an integral over space. This will be a major difference with respect to the non-projectable case.

We now move straight to the linearization. If for simplicity one sets the cosmological constant term to 0, then flat space is a suitable background. After suitable gauge choices, see Ref.~\cite{Sotiriou:2009bx}, one finds that the theory propagates a spin-2 mode, the usual graviton, that now satisfies a modified dispersion relation, {\em i.e.}
\begin{equation}
\ddot{ \widetilde H}_{ij}   =  -\left[  g_1  \partial^2   + g_3  M_{\rm pl}^{-2} \partial^4   + g _8  M_{\rm pl}^{-4} \partial^6 \right] \widetilde H_{ij},
\end{equation}
where ${\widetilde H}_{ij}$ is transverse and traceless. Notice that the couplings $g_3$ and $g_8$ control the scale that the lorentz violating terms become important in the dispersion relation. Recall also that in vacuum one has the freedom to rescale the coordinates and set $g_1=-1$.

As mentioned earlier, less symmetry generically means more degrees of freedom, so since the theory is invariant under foliation preserving diffeomorphisms only, we expect to find extra excitations. This is indeed the case. There is an extra scalar degree of freedom, whose linearized dynamics are governed by the action \cite{Sotiriou:2009bx}
\be
S^p_2=- M_{\rm pl}^2 \int  d^3x d t  \left[{1 \over c_h^2}\dot{h}^2 + h\partial^2 h+\frac{8g_2+3g_3}{M_{\rm pl}^2} h\partial^4 h -\frac{8 g_7-3 g_8}{M_{\rm pl}^4} h \partial^6 h  \right ] , \label{quad}
\ee
where
\be
c_h^2=\frac{1-\lambda}{3 \lambda -1},
\ee
and we have set $g_1=-1$ by rescaling the coordinates, in order to guarantee the stability of the spin 2 graviton.
Given the overall minus sign in eq.~(\ref{quad}), the scalar mode is a ghost whenever $1>\lambda>1/3$. On the other hand, the scalar is classically unstable at low energies whenever $\lambda>1$ or $\lambda<1/3$ \cite{Sotiriou:2009bx}.  It is conceivable that the instability can be cut off by the higher order derivatives. If we call $M_\star$ the scale where the higher order term would take over, then the time scale of the instability would be at least $1/|c_h| M_\star$. The relation between $M_\star$ and $M_{\rm pl}$ is clearly governed by the $g_i$ coefficients in eq.~(\ref{quad}). On the other hand, if the instability is not to develop within the lifetime of the universe then $|c_h| M_\star< H_0$, where $H_0$ is the Hubble parameter.  \cite{Koyama:2009hc}. Experiments rule out modifications of Newton's law at scales above $10\mu m$ which corresponds roughly to the constraint $M_\star\gtrsim 0.1 {\rm eV}$. In turn, this implies $|1-\lambda|\lesssim 10^{-61}$. This value is clearly very low and it seems unlikely that the renormalization group flow could drive $\lambda$ to it.

In Refs.~\cite{Huang:2010rq,Wang:2010ug} linearization around a de Sitter, instead of Minkowski, background was considered. As expected, the large scale modes exhibit better behaviour. However, qualitatively the result is the same overall: $\lambda$ needs to be sufficiently close to $1$ for the instability not to be a concern. Note that requiring $\lambda$ to be very close to one is not only a potential fine tuning problem. $c_h\rightarrow 0$ as $\lambda\rightarrow 1$, which means that the low momentum phase velocity of the scalar mode will be much smaller than the speed of light for such values of $\lambda$. When a velocity of a certain mode is smaller than the speed of light in a medium, relativistic particles traveling in this medium decay via the Cherenkov process \cite{Elliott:2005va}. The fact that we observe high energy cosmic rays which need to travel a significant distance to reach us is evidence enough that such modes do not exist.

\subsection{Strong coupling}

In the previous section we only considered the linearized dynamics as described by the part of the perturbative action quadratic in $h$. However, it has been pointed out by several authors that such a perturbative treatment breaks down when $\lambda$ approaches $1$ and the scalar mode gets strongly coupled \cite{Charmousis:2009tc,Blas:2009yd,Koyama:2009hc}. To see how this comes about we give below the cubic interactions of $h$,
\be
S^p_3=M_{\rm pl}^2\int d t d^3x \left\{h (\partial h)^2  - {2 \over c_h^4} \dot{h} \partial_i h {\partial^i \over \partial^2} \dot{h}   
+  \frac{3}{2}\left[{1 \over  c_h^4}    h \left( {\partial_i \partial_j \over  \partial^2}\dot{h}\right)^2 - {(2 c_h^2+1)  \over  c_h^4}  h \dot{h}^2   \right]
  \right\} \label{cubic},
  \ee
where we have neglected cubic interactions coming from higher order operators. The redefinitions $\hat{t}=|c_h| t$ and $\hat{h}=c_h^{-1/2} M_{\rm pl}\,  h$ canonically normalize the lower operator part of the quadratic action (\ref{quad}) (first two terms). After these redefinitions the cubic action reads
\be
S^p_3=\frac{1}{|c_h|^{3/2} M_{\rm pl}}\int d \hat{t} d^3x \left\{
c_h^2\,\hat{h} (\partial \hat{h})^2  - 2 \hat{h}' \partial_i \hat{h} {\partial^i \over \partial^2}\hat{h}'   
+  \frac{3}{2}\left[   \hat{h} \left( {\partial_i \partial_j \over  \partial^2}\hat{h}'\right)^2 - (2 c_h^2+1)  \hat{h}(\hat{h}')^2   \right]
  \right\} \label{cubicred},
  \ee
  where now $'=\partial/\partial_{\hat{t}}$. As is obvious from the equation, there are cubic interactions which are suppressed by the scale $|c_h|^{3/2} M_{\rm pl}$ with respect to the quadratic ones. Therefore, the theory gets strongly coupled at the scale $M_{sc}=|c_h|^{3/2} M_{\rm pl}$.

Given the constraints on $\lambda$ from stability discussed previously, the strong coupling scale is phenomenologically unacceptably low, as we know that we can treat gravity perturbatively at low energies. There is also the issue of renormalizability: our arguments for power-counting renormalizability given above where based on the validity of the perturbative treatment at all energies. If there is strong coupling such arguments simply fail. 
 
The only subtlety here is the fact that the strong coupling arguments presented so far neglect the higher order operators. However, for such operators to become important and invalidate our treatment, the strong coupling scale $M_{sc}$ would have to be higher than the scale $M_\star$ that suppresses them. Given how low $M_{sc}$ is here, this would imply that higher order operators modify the graviton dynamics at very low energies, which would be in conflict with current experimental evidence.

Since the analysis presented here is linear, one could ask whether including non-linear effects could resolve the strong coupling problem, by an analogue of the Vainstein mechanism in massive gravity \cite{vainshtein}, {\em i.e.}~a non-perturbative restoration of  the $\lambda\rightarrow 1$ limit. Recently, in Ref.~\cite{Mukohyama:2010xz} it has been claimed that this is indeed the case in spherically symmetric, static configurations. This matter deserves further attention.

\section{Non-projectable version: Dynamics, consistency and low energy behaviour}
\subsection{Degrees of freedom and propagators}

We now turn our attention to the non-projectable version of the theory. As before, we will not enforce detailed balance in order to obtain the most general results possible. However, we will discuss explicitly that case as well.
The dynamics of the spin 2 graviton are essentially the same as in the projectable version so we will not consider them again in more detail. 
As in the projectable case, also here there is an extra scalar degree of freedom, and we will focus on that. 

The low energy dynamics of this scalar graviton are governed by the action
\be
S^{np}_2=- M_{\rm pl}^2 \int  d^3x d t  \left[{1 \over c_h^2}\dot{h}^2 + \frac{\eta-2}{\eta} h\partial^2 h  \right ] , \label{quadnp}
\ee
which one obtains by linearizing action (\ref{npaction}) around flat space and considering only the lowest order operators. $\xi$ has been set to $1$ by a suitable coordinate rescaling. Note that $c_h$ is defined in the same way as above and, thus, it is {\em not} the low momentum phase velocity of the scalar in the non-projectable version. The square of the latter is now
\be
c^{\prime 2}_h= c^2_h\frac{\eta-2}{\eta}.
\ee
For $h$ not to be a ghost one needs $c_h^2<1$, and for $h$ to be classically stable one needs $c_{h}^{\prime 2}>0$. These conditions are satisfied whenever \cite{Blas:2009qj}
\be
\label{npcon}
\lambda>1, \quad 0<\eta<2,
\ee
or
\be
\lambda<1/3, \quad 0<\eta<2.
\ee
In the second region in the parameter space $\lambda$ is far from $1$, which is the value it has in general relativity, so we will not consider this option further. 
The presence of the operator $a_i a^i$ in action (\ref{npaction}), which is a lowest order operator and contributes to the quadratic action, has drastically improved the behaviour of the scalar graviton with respect to the projectable case we examined previously. In particular, there is now a region in parameter space, described by the constraints (\ref{npcon}), for which $h$ is neither a ghost nor classically unstable \cite{Blas:2009qj}.  Even though we have neglected higher order operators here, the behaviour we have found would remain qualitatively the same if we had included them. 

Let us examine what would happen if we had enforced detailed balance. The $a^i a_i$ term in the action would not be allowed which means $\eta=0$. In this case, the coefficient of the second term in (\ref{quadnp}) blows up and the scalar field appears to freeze. Things, however, are actually more complicated, as a more detailed analysis shows \cite{Blas:2009yd}. When detailed balance is enforced, the scalar does propagate around backgrounds that are neither static nor homogeneous. In fact it appears to satisfy a first order differential equation, which constitutes a serious problem for the theory, see Ref.~\cite{Blas:2009yd} for more details. Similar conclusions have been obtained by considering the hamiltonian formulation of the theory with detailed balance \cite{Li:2009bg,Henneaux:2009zb}. In particular, the Hamiltonian constraint is not automatically preserved by time evolution (second class constraint) and the theory propagate $5/2$ half degrees of freedom, the $1/2$ corresponding to the scalar mode which satisfies a first order equation and the rest to the two graviton polarizations. Such a theory, though mathematically consistent, exhibits no time evolution and is, therefore, physically meaningless \cite{Henneaux:2009zb}.

\subsection{Strong coupling}

We just saw that, judging from the quadratic action, the non-projectable version exhibits improved scalar dynamics with respect to the projectable version, as long as detailed balance is not enforced. However, we found earlier that the scalar mode in the projectable version gets strongly coupled at unacceptably low energies, so one has to check whether this is the case in the non-projectable version as well. To do so, we write down the cubic action for the scalar \cite{Papazoglou:2009fj}:
\bea
S^{np}_3&=&M_{\rm pl}^2\int d t d^3x \left\{\left(1-\frac{4(1-\eta)}{\eta^2}\right)h (\partial h)^2  - {2 \over c_h^4} \dot{h} \partial_i h {\partial^i \over \partial^2} \dot{h}     \right. \nonumber \\&& \left.~~~~~~~~~~~~~~~~~~~ 
+  \left(\frac{3}{2}+\frac{1}{\eta}\right)\left[{1 \over  c_h^4}    h \left( {\partial_i \partial_j \over  \partial^2}\dot{h}\right)^2 - {(2 c_h^2+1)  \over  c_h^4}  h \dot{h}^2   \right]
  \right\} \label{cubicnp}.
  \eea
Same as before for the projectable case, we canonically normalize the low energy quadratic action. This requires the redefinitions
\be
t=\sqrt{\frac{\eta}{2-\eta}}\frac{\hat{t}}{|c_h|}, \qquad h=\left(\frac{\eta}{2-\eta}\right)^{1/4} \sqrt{|c_h|} \frac{\hat{h}}{M_{\rm pl}}.
\ee 
 Recall that we are only interested in the region of the parameter space where conditions (\ref{npcon}) are satisfied. Under the redefinitions the cubic action reads
\bea
S^{np}_3&=&\frac{(2-\eta)^2}{\eta^{1/2} c^{\prime\,3/2}_h M_{\rm pl}}\int d \hat{t} d^3x \left\{
c_h^2\,\left(1-\frac{8(1-\eta)}{(2-\eta)^2}\right)\hat{h} (\partial \hat{h})^2  - 2 \hat{h}' \partial_i \hat{h} {\partial^i \over \partial^2}\hat{h}'     \right. \nonumber \\&& \left.~~~~~~~~~~~~~~~~~~~ 
+  \left(\frac{3}{2}+\frac{1}{\eta}\right)\left[   \hat{h} \left( {\partial_i \partial_j \over  \partial^2}\hat{h}'\right)^2 - \left(\frac{2\, \eta\, c_h^{\prime\,2}}{2-\eta}+1\right)  \hat{h}(\hat{h}')^2   \right]
  \right\} \label{cubicrednp},
  \eea
  where we have expressed everywhere $c_h$ in terms of the physical quantity $c'_h$. From eq.~(\ref{cubicrednp}) one directly infers that the cubic interactions are suppressed with respect to the quadratic ones by various scales $f(|\lambda-1|,\eta) M_{\rm pl}$, where $f$ is an algebraic function whose functional form depends on which terms one considers. The scalar mode would become strongly coupled at the lowest of these scales.
  
Both $|\lambda-1|$ and $\eta$ essentially measure deviations from Lorentz invariance which are present at arbitrarily low scales. They are to be small for the theory to avoid experimental constraints on Lorentz violations. It is rather straightforward to verify that in this case, also the scale that the theory gets strongly coupled has to be low. Instead of performing a general analysis leaving both $|\lambda-1|$ and $\eta$ small but arbitrary, we will focus on the most interesting case where $c'_h\sim 1$. This value of $c'_h$ is the one dictated by Cherenkov radiation constraints  \cite{Elliott:2005va}, see discussion above. We then have $\eta\sim|\lambda-1|$ and the strong coupling scale turns out to be $M_{sc}\sim\sqrt{\eta} M_{\rm pl}\sim\sqrt{|\lambda-1|} M_{\rm pl}$ \cite{Papazoglou:2009fj,Kimpton:2010xi}.
  
Since $\eta, |\lambda-1|\ll 1$ the strong coupling scale will be much lower than the Planck scale. Absence of preferred frame effects in the Solar system observations requires $\eta, |\lambda-1| \lesssim 10^{-7}$ \cite{Will:2005va}, which in turn implies that $M_{sc} \lesssim 10^{15} {\rm GeV}$. Therefore, the strong coupling scale is too high to be phenomenologically accessible from gravitational experiments. In this sense, strong coupling in non-projectable Ho\v{r}ava--Lifshitz gravity would pose no threat if the theory were to be treated as an effective theory describing gravity at low energies. On the other hand, the mere presence of strong coupling even at relatively high energies, as in this case, is enough to seriously question the power counting renormalizability of the theory \cite{Papazoglou:2009fj}. Indeed, as mentioned also earlier when considering the projectable case, renormalizability arguments were based on the assumption that a perturbative treatment can be used to arbitrarily high energies.

However, the strong coupling arguments given here neglect the role of the higher order operators in the action. As proposed in Ref.~\cite{Blas:2009ck}, a way to avoid strong coupling is to lower the scale that suppresses the higher order operators in action (\ref{npaction}) below $M_{sc}$, {\em i.e.}~impose that $M_A\sim M_B\sim M_\star$ and $M_\star<M_{sc}$. This clearly requires the introduction of a second scale, different from $M_{\rm pl}$, as well as a peculiar hierarchy of scales. Also, it does seem to involve some large dimensionless couplings as $M_{\rm pl} \gg M_{\star}$. However, as it has been argued in Refs.~\cite{Blas:2009ck,Blas:2010hb} this hierarchy of scales is technically natural.

An important consequence of this effective constraint on $M_\star$ coming from the requirement to avoid strong coupling, first pointed out in Ref.~\cite{Papazoglou:2009fj}, is that now $M_\star$ becomes bound from both above and below. Being the scale that suppresses higher order derivatives in the dispersion relations, one would expect that the larger $M_\star$ is, the better. Assuming that this same scale suppresses higher order derivatives in the matter dispersion relations (which would also be modified in a way similar to gravity), and in particular photons, places a strong lower bound on $M_\star$: The absence of a time delay in different frequency $\gamma$-rays coming from distant astrophysical sources implies that \cite{Albert:2007qk,:2009zq} $M_\star \gtrsim 10^{11}{\rm GeV}$. However, now one has also an upper bound if strong coupling is not to be a problem, and the combined set of constraints yields
\be
10^{15} {\rm GeV} \gtrsim M_\star \gtrsim 10^{11}{\rm GeV}.
\ee 
There seems to be a comfortable window for $M_\star$ within which non-projectable Ho\v{r}ava--Lifshitz gravity (without detailed balance or other restrictions) avoids strong coupling without exhibiting detectable Lorentz violations, at least with current experimental accuracy. Improving the accuracy with which we can constrain both preferred frame effects in the solar system  and modifications in the dispersion relations could potentially close this window.
 
\subsection{Relation to Einstein-Aether theory}

Non-projectable Ho\v{r}ava--Lifshitz gravity is a theory with a preferred foliation, which will exhibit Lorentz invariance violations even at low energies. This should be clear by now by inspecting the action (\ref{npaction}) and taking into account that, given the dynamics of the scalar discussed previously, one can no longer hope that $\lambda$ and $\eta$ will run to $1$ and $0$ respectively in the IR, in order for Lorentz symmetry to emerge. Note that the operators related to these couplings are of lowest order, {\em i.e.}~dimension 2. 

A model theory for gravity with preferred frame effects is Einstein-aether theory \cite{Jacobson:2000xp,Jacobson:2008aj}. It is the most general theory for a unit timelike vector field coupled to gravity (but not to matter), which is second order in derivatives. 
The most general action for Einstein-aether theory, up to total derivative terms and setting aside matter coupling, is
\be \label{S}
S_{\ae} = \frac{1}{16\pi G_{\ae}}\int \sqrt{-\bar{g}}~ (-\bar{R} + L_{\ae})
~d^{4}x 
\ee
where $\bar{g}_{\mu\nu}$ is the 4-dimensional metric, $\bar{g}$ its determinant, $\bar{R}$ the Ricci scalar of this metric, $\bar{\nabla}_\mu$ is the covariant derivative associated with $\bar{g}_{\mu\nu}$,
\be \label{Lae}
L_{\rm \ae} = -M^{\a\b\m\n} \bar{\nabla}_\a u_\m \bar{\nabla}_\b u_\n, 
\ee
with $M^{\a\b\m\n}$ defined as
\bea M^{\a\b\m\n} = c_1 \bar{g}^{\a\b}\bar{g}^{\m\n}+c_2\bar{g}^{\a\m}\bar{g}^{\b\n}
+c_3 \bar{g}^{\a\n}\bar{g}^{\b\m}+c_4 u^\a u^\b \bar{g}_{\m\n}. 
\eea
Greek indices run from $0$ to $3$, the $c_i$ are dimensionless coupling constants,
and it is assumed that $u_\m$ is constrained to be a unit 
vector, $g^{\m\n}u_\m u_\n=1$. This constraint can be explicitly imposed by the use of a lagrange multiplier, or implicitly taken into account in the variation of the action by allowing only variations that respect it. 

One may wonder whether there is any relation between Einstein-aether theory and the (naive) IR limit of non-projectable Ho\v{r}ava--Lifshitz gravity (that is, a theory described by action (\ref{npaction}) without the higher order operators suppressed by $M_A$ and $M_B$). It was shown in Ref.~\cite{Jacobson:2010mx} that the latter is actually a limiting case of the former. Let us briefly go through the arguments presented there.

Starting from the action (\ref{S}), impose the restriction that the aether be hypersurface orthogonal:
\be
\label{ho}
u_\a=\frac{\partial_\a T}{\sqrt{g^{\m\n}\partial_\m T \partial_\n T}}\,,
\ee 
where the unit constraint on the aether has been taken into account. $T$  will satisfy a dynamical equation. However, since it is a scalar, on shell its dynamical equation will be implied by the contracted Bianchi identity, provided that $T$ is not a constant (this can be shown for a general scalar field coupled to gravity, provided that the theory is manifestly covariant and the other fields are assumed to satisfy their field equations). However, $T$ cannot be a constant as it now defines the foliation. Thus, we are allowed to not explicitly impose the equation of motion for $T$, which in turn implies that $T$ itself can be chosen as the time coordinate. We then have
\be
\label{ut}
u_\a=\delta_{\a T} (g^{TT})^{-1/2}=N\delta_{\a T}  \,.
\ee
This choice selects a preferred foliation with $N$ being the lapse function of this foliation. Substituting eq.~(\ref{ut}) into the Einstein-aether theory action (\ref{S}), one gets after some algebra
\be\label{SBPSH}
S= \frac{M_{\rm pl}^2}{2}\!\int dt d^3x \, N\sqrt{g}(K_{ij}K^{ij} - \lambda K^2 
+ \xi {}^{(3)}\!R + \eta\, a_ia^i),
\ee
with the following correspondence of parameters: 
\be
\label{HLpar}
\frac{1}{8\pi M_{\rm pl}^2 \,G_{\ae}}=\xi=\frac{1}{1-c_{13}}, \quad \lambda=\frac{1+c_2}{1-c_{13}},\quad \eta=\frac{c_{14}}{1-c_{13}},
\ee
where we use the convention $c_{ij}=c_i+c_j$.\footnote{Note that there is a typo in the formula relating $\xi$ and $c_{13}$ in eq.~(10) of Ref.~\cite{Jacobson:2010mx}.} Clearly, action (\ref{SBPSH}) is the same as action (\ref{npaction}) without the higher order operators. Therefore, the IR limit of non-projectable Ho\v{r}ava-Lifshitz gravity is equivalent to Einstein-aether theory with a hypersurface orthogonal aether. 

This equivalence provides new insights to both theories. For Einstein-aether theory it implies that there may exist a UV completion, at least for some subclasses. For non-projectable Ho\v{r}ava--Lifshitz gravity it tells us that it can be viewed as a theory were a preferred foliation is dynamically defined by a scalar field. Note that one would expect to be able to supplement Einstein-aether gravity (with a hypersurface orthogonal aether) with suitable higher order operators and construct a covariant equivalent of the full non-projectable Ho\v{r}ava--Lifshitz gravity (see Ref.~\cite{Germani:2009yt} for an early attempt).

Clearly, on account of the equivalence,  some results obtained for Einstein-aether theory go through to the IR limit of Ho\v{r}ava--Lifshitz gravity and vice versa. A characteristic example is spherically symmetric solutions, as the aether is always hypersurface orthogonal in spherical symmetry \cite{Jacobson:2010mx,Blas:2010hb}. For example, some spherically symmetric solutions found in Ref.~\cite{Kiritsis:2009vz} for Ho\v{r}ava--Lifshitz gravity had actually already been found earlier for Einstein-aether theory \cite{Eling:2006df}. On the other hand, some important results do not go through from general Einstein-aether theory, as they hinge on the exact form of the field equations, and specifically on whether the vector is a gradient of a scalar or not. For instance, though qualitatively similar, quantitatively the PPN constraints on non-projectable Ho\v{r}ava--Lifshitz gravity cannot simply be inferred from those of Einstein-aether theory via a parameter mapping and have been obtain independently in Ref.~\cite{Blas:2010hb}.

\section{Cosmology}

As already mentioned, all versions of Ho\v{r}ava--Lifshitz gravity presented above are written in a preferred foliation, and, therefore, they do not constitute diffeomorphism invariant theories, but they are invariant under diffeomorphisms that preserve the foliation. This symmetry provides enough gauge freedom to choose
\be
N=1,\quad N^i=0,\quad g_{ij}=a(t)^2\delta_{ij},
\ee
and bring the ADM line element (\ref{admmetric}) into the usual Friedmann--Lema\^{i}tre--Robertson--Walker (FLRW) form
\be
ds^2=-dt^2+a(t)^2 \left[\frac{dr^2}{1-k r^2}+r^2\left(d\theta^2+\sin^2 \theta d\phi^2\right)\right],
\ee
under the usual cosmological assumptions of homogeneity and isotropy. Here, $a(t)$ is the scale factor and $k=+1, 0, -1$ according to whether the universe is hyperspherical, spatially flat, or hyperbolic.

Clearly, in this setting $N$ has no space dependence. Therefore, its spatial derivatives vanish. This will make the actions of the projectable and the non-projectable versions of the theory, eqs. (\ref{paction}) and (\ref{npaction}) coincide (assuming that all operators allowed by the symmetries of the theory have been consistently taken into account). The field equations do not coincide completely, as it does make a difference whether $N$ is assumed to be space-independent {\em a priori} or whether the latter is enforced as a gauge choice. The difference lies on the Hamiltonian constraint and the rest of the equations coincide. The former is global in the projectable theory and local in the non-projectable theory. So, when it comes to background cosmology, one can study both version of the theory at the same time, just taking into account the subtle issue of the Hamiltonian constraint. The dynamics of perturbation around the background will, of course, differ in the two versions, but we will not discuss perturbations here.

Cosmology in Ho\v{r}ava--Lifshitz gravity was first studied in Ref.~\cite{Kiritsis:2009sh,Sotiriou:2009bx} (with detailed balance in Ref.~\cite{Calcagni:2009ar}). A recent review of cosmology in the projectable version has recently been given in Ref.~\cite{Mukohyama:2010xz}, and as explained above, most of the results summarized there are applicable also on the non-projectable version. Here we will only give a very brief overview of background cosmology and mention in passing some key advantages of Ho\v{r}ava--Lifshitz cosmology. We refer the reader to the relevant literature for more details.

We start by presenting the field equations in an FLRW background. The supermomentum constraint is trivially satisfied, so we are left with two equations, corresponding to the first and second Friedmann equations. In the projectable version, the Hamiltonian constraint yields
\begin{equation}
\label{f1p}
\int d^3 x a^3 \left\{\frac{3\lambda-1}{2}\; {\dot a^2\over a^2}  -  {V(a)\over 6}  - {8\pi G_N \rho\over 3}\right\}=0,
\end{equation}
where $8\pi G_N\equiv M_{\rm pl}^{-2}$, $\rho\equiv-g^{-1/2} \delta S_M/\delta N$,  $S_M$ is the matter action, and
\begin{eqnarray}
V(a) = g_0\, M_{\rm pl}^2 +  {6 g_1 k  \over  a^2} + {12(3g_2+g_3)  k^2 \over M_{\rm pl}^2\,a^4} 
+ {24(9 g_4 + 3g_5+ g_6) k\over M_{\rm pl}^4 \, a^6}. 
\end{eqnarray}
In the non-projectable case, the Hamiltonian constraint is local and one can do away with the integral:
\begin{equation}
\label{f1np}
\frac{3\lambda-1}{2}\; {\dot a^2\over a^2}  -  {V(a)\over 6}  = {8\pi G_N \rho\over 3}.
\end{equation}
The dynamical equations yield in both cases
\begin{eqnarray}
\label{f2}
- \frac{3\lambda-1}{2} \; {\ddot a\over a}   &=&   {1\over2}\frac{3\lambda-1}{2} {\dot a^2\over a^2}   
-  {1\over12 a^2} {d[V(a)\, a^3]\over d a}
+ 4\pi G_N p,
\end{eqnarray}
where $p\equiv -g^{ij}(2/N\sqrt{g})\delta S_m/\delta g^{ij}$.

Eqs.~(\ref{f1p}) and (\ref{f2}) and Eqs.~(\ref{f1np}) and (\ref{f2}) are governing background cosmology in the projectable and non-projectable versions respectively. Notice the following: In the non-projectable version one can eliminate $\dot{a}$ from eq.~(\ref{f2}) by using eq.~(\ref{f1np}) to get
\begin{eqnarray}
\label{f3np}
- \frac{3\lambda-1}{2} \; {\ddot a\over a}   &=&    -{1\over 12 a} {d[V(a) a^2] \over d a}  +{4\pi G_N\over 3} (\rho+3p).
\end{eqnarray}
Differentiating eq.~(\ref{f1np}) and subtracting from eq.~(\ref{f3np}) yields the standard conservation law
\be
\label{mc}
\dot{\rho}+3\frac{\dot{a}}{a} (\rho+p)=0.
\ee
This implies that the conservation law for matter is implied by the two (modified) Friedmann equations and, thus, imposing it separately is neither needed nor will it constrain the dynamics. This is in direct analogy to general relativity.

Things are quite different in the projectable case, given the global nature of the Hamiltonian constraint. It has been argued in Ref.~\cite{Mukohyama:2009mz} that a global constraint, such as eq.~({\ref{f1p}), is irrelevant for the local patch of the universe inside the Hubble horizon which the FLRW spacetime is supposed to approximate. Now, we could, instead of differentiating eq.~(\ref{f1p}), integrate eq.~(\ref{f2}) and ignore eq.~(\ref{f1p}) completely. One then gets
\begin{equation}
\label{f1p2}
\frac{3\lambda-1}{2}\; {\dot a^2\over a^2}  -  {V(a)\over 6}  = \frac{8\pi G_N}{3} \left(\rho+\frac{C(t)}{a^3}\right),
\end{equation}
where the functional form of $C(t)$ depends on the conservation law matter satisfies. If we eliminate $\dot{a}$ from eq.~(\ref{f2}) by using eq.~(\ref{f1p2}) we get
\begin{eqnarray}
\label{f3p}
- \frac{3\lambda-1}{2} \; {\ddot a\over a}   &=&    -{1\over 12 a} {d[V(a) a^2] \over d a}  +{4\pi G_N\over 3} \left(\rho+\frac{C(t)}{a^3}+3p\right).
\end{eqnarray}
Now, the conservation law for matter has become essential to the dynamics via the presence of $C(t)$. The terms containing $C(t)$ are the only difference between eqs.~(\ref{f1p2}) and (\ref{f3p}), which describe the background dynamics in the projectable case, and eqs.~(\ref{f1np}) and (\ref{f3np}), which describe the background dynamics in the non-projectable case.\footnote{This analysis, leading to the same conclusions, can be performed also before imposing the FLRW ansatz and the relevant symmetries, see Ref.~\cite{Mukohyama:2009mz}.}

Now that we have the field equations in a suitable form for both cases we can proceed to discuss phenomenology. Suppose that matter is coupled in such a way that at low energies $\rho$ and $p$ have the usual meaning and satisfy the standard conservation law of eq.~(\ref{mc}) (this is a strong assumption, see also next section). For the projectable case this implies that $C(t)=C_0$, where $C_0$ is a constant. The $C_0$ related terms then play the role of a dark matter component \cite{Mukohyama:2009mz}. This dark matter component will not exist in the non-projectable case.

Setting aside the role of matter, the main difference between Ho\v{r}ava--Lifshitz gravity and general relativity in background cosmology is the presence of the last two terms in $V$. The first one is what is known as a dark radiation component, as it scales as $a^{-4}$. The second one is what one could call a stiff matter component, as it scales as $a^{-6}$. Note that the sign of the $a^{-6}$ term depends on the sign of the $k$ and that both terms vanish when $k=0$. An interesting feature related to the presence of these terms is that they can lead to solutions with classical cosmological bounces, provided that their coefficients will have appropriate signs \cite{Sotiriou:2009bx,Kiritsis:2009sh}.

One last important property of Ho\v{r}ava--Lifshitz cosmology which is worth mentioning before closing this brief overview, is that is seems to lead to a scale invariant spectrum of cosmological perturbations, without the need for inflation \cite{Mukohyama:2009gg,Calcagni:2009qw}. This property is related to the anisotropic scaling between space and time that the theory exhibits at high energies, which is also the key ingredient for power counting renormalizability.

\section{Conclusions and future perspectives}

A brief overview of Ho\v{r}ava--Lifshitz gravity was given and special attention was paid in distinguishing between the various different versions. After that, the main focus of this review was on the dynamics and the viability of these versions. In summary, the projectable version has serious viability problems because it harbours a scalar mode which is either classically or quantum mechanically unstable (depending on the position in parameter space) and also exhibits strong coupling at low energies (whether or not the Vainshtein mechanism can alleviate the latter is not clear yet, see also above). The non-projectable version, on the other hand, seems to have similar (and other) problems if detailed balance is imposed. These problems, however, appear to be overcome once detailed balanced is abandoned and all operators allowed by the symmetries of the theory are consistently taken into account. Alleviating the strong coupling problem requires the introduction of a second scale in the theory which should be parametrically smaller than the Planck scale, and should be the scale at which higher order operators become important. The dynamics of the version exhibiting the extra $U(1)$ symmetry, which was very recently proposed, have not been adequately studied yet.

There are many open issues regarding Ho\v{r}ava--Lifshitz gravity, the two most important ones being the following. Firstly, in the introduction we argued that the theory is power-counting renormalizable, as is usually done in the literature. Even though this is a strong indication for UV completeness,  renormalizability beyond power counting has not been explicitly shown (see Ref.~\cite{Orlando:2009en} for some work in this direction). Additionally, the renormalization group flow of the various couplings has not been studied, which implies that, for the time being, one does not really know if the theory approaches GR in the IR ($\lambda\to 1$, $\eta\to 0$) or not. Secondly, the role of matter and its coupling to gravity have not been fully clarified yet. The matter action will have to include higher order spatial derivatives, which implies that there will be modifications in the dispersion relations of matter fields that can lead to serious constraints. Additionally, couplings between the matter and the scalar graviton could lead to violations of the equivalence principle. Even if initially omitted, such couplings would be generically generated by radiative corrections, unless some symmetry prohibits them ({\em e.g.}~supersymmetry \cite{GrootNibbelink:2004za}). Hopefully, future work will shed some light onto these issues.

\ack

I am indebted to my collaborators Antonios Papazoglou, Matt Visser and Silke Weinfurtner  for their hard work and invaluable input in our joint projects on Ho\v{r}ava--Lifshitz gravity. I would also like to thank Ted Jacobson, Ian Kimpton and Tony Padilla for numerous stimulating and enlightening discussions. Many thanks also to the participants of the Peyresq 2010 meeting. This work was supported by a Marie Curie Incoming International Fellowship.

\end{document}